# Optimal Bitter Coil Solenoid

V. **Kobelev**

**Abstract.** For generation of extremely strong magnetic fields the electromagnet in form of the Bitter coil is used. A number of factors limit the upper bound of magnet flux density. The high stresses due to Lorentz forces in the coil is one principal constraint. The Lorentz force acts as the pressure of magnetic field. The Lorentz forces generate the distributed body force. The common radial thickness profile of the Bitter coil is constant. In this paper the possibility of optimization by means of non-constant radial thickness profile of the Bitter coil is studied. The close form expression for optimal thickness profile is obtained. Both designs – the constant one and the optimally shaped - are compared.  The considerable improvement of magnetic flux density due to shape optimization is demonstrated. Moreover, the optimal design improves the shape of cooling channels. Namely, the highest cross-section of cooling channel is at the most thermally loaded inner surface of the coil.

# 1. Introduction

The problem of solenoid stress and strain has been addressed over an extended period of time with increasing generality. The major developments in the evolution of an analytical treatment are reviewed here, relating assumptions and simplifications made by various authors.

The Bitter type magnet consists of a sequence of annular conducting plates (Herlach, Jones, 1994). Each plate of posesses a radial cut, resulting in radial current distribution distribution. By pressure contact between two sectors adjacent to the cut the series electrical connections between plates are performed. The remaining flat contact area is insulated by insertion of a insulator material. Bitter magnets are the frequently used in the magnet installations. A more recent development is the polyhelix design

The present work is based on the common assumptions (Lontai and Marston, 1965). The coils were assumed to be a homogeneous, linear elastic material. The current density leads to the distributed Lorentz body force. A stress analysis of superconducting solenoids was presented (Markiewicz et al, 1999) which includes a generalized plane strain condition for the axial strain.

The design of pulsed coils for excitation of high magnetic fields is considered in the book (Kratz, Wyder, 2013)

## 2. STRESS BALANCE EQUATIONS FOR MAGNETIC SOLENOID, GENERAL FORM

The solenoid winding is assumed to be axial symmetric. The inside and outside radii of the coil are

$$R_1 < r < R_2, \quad R_2/R_1 = \mathrm{K} > 1$$

Due to the symmetry of winding, the current in the coil, the Lorentz force density, the mechanical stresses and the deformation on the coil are all axial symmetric as well.

For formulation of the governing equations of the solenoid the cylindrical coordinates are used. The general displacement of a point in a cylinder is described by its components along the coordinate axes. Generally saying, due to symmetry the displacement components are the functions of radius and axial coordinate:

$$u(r,z), \qquad v(r,z) = 0, \qquad w(r,z) .$$

For simplicity the axial deformation is ignored, such that the displacements are assumed to be the following functions of radius only:

$$u(r), \qquad v = 0, \qquad w = 0 .$$

With this assumption the nonzero components of strain are:

$$\varepsilon_r = \frac{du}{dr}, \qquad \varepsilon_\theta = \frac{u}{r} \tag{2.1}$$

The mechanical strain is the sum of the strain due the Lorentz forces and strain due to thermal contraction loads. The axial contraction stress is neglected:

$$\sigma_z = 0 ,$$

because the analysis is performed for a single coil. For simplification the thermal contraction loads are also not considered.

The constitutive equations of linear elastic isotropic material lead to the expressions of stresses. With the above assumptions for a magnetic solenoid, the only nonzero components of stress are

$$\sigma_r = \frac{E}{1-v^2}\left(\frac{du}{dr} + v\frac{u}{r}\right), \qquad \sigma_\theta = \frac{E}{1-v^2}\left(v\frac{du}{dr} + \frac{u}{r}\right). \tag{2.2}$$

The equivalent stress in the material is restricted

$$\sigma_{eqv} = \frac{1}{\sqrt{2}}\sqrt{(\sigma_r - \sigma_\theta)^2 + \sigma_r^{\ 2} + \sigma_\theta^{\ 2}} \le \sigma_0 . \tag{2.3}$$

The thickness $h(r)$ of the Bitter plate is assumed to be a function of radius. The general stress balance equation for a body of a variable thickness $h(r)$ with a distributed force $F(r)$ reads

$$\frac{d[h(r)\sigma_r]}{dr} + h(r)\frac{\sigma_r - \sigma_\theta}{r} + F = 0 . \tag{2.4}$$

The distributed Lorentz force density per unit radius of the windings reads:

$$F = B(r)h(r)j(r). \tag{2.5}$$

The magnetic field intensity on the axis of the coil is:

$$B(r) = \int_r^{R_2} h(r)B_j(r)dr, \qquad B_j(r) = \frac{\mu_0 j(r)}{2r}, \qquad \mu_0 = \frac{4\pi}{10^7}. \tag{2.6}$$

The total current through one Bitter coil is:

$$J_0 = \int_{R_1}^{R_2} h(r)j(r)dr, \tag{2.7}$$

where $j(r)$ is the current density pro unit area.

The value of the total current $J_0$ could be arbitrary. Obviously, the higher this value is, the greater is value magnetic field intensity in the inside Bitter plate. However, the upper value of total current is strongly limited by different factors. The most important factors are the conductive heating of coil and the stresses in the coil due to Lorenz pressure of magnetic field. In this article the only limitation is the admissible tension stresses due to pressure of stored magnetic field.

The current density distribution guarantees the constant electric potential over all contours with different radii $r$ is:

$$j(r) = \frac{a}{r}. \tag{2.8}$$

The constant $a$ depends on the actual thickness function $h(r)$. If the thickness function is given, the value of constant $a$ determines form Eq. (2.7).

The volume of the conducting material reads:

$$V = 2\pi \int_{R_1}^{R_2} h(r)rdr. \tag{2.9}$$

## 3. Static magnetic analysis of the Bitter coil of constant thickness

At first consider the Bitter coil of the constant thickness:

$$h(r) = h_c .$$

Our next task is to determine the thickness $h_c$, assuming the total current in the coil $J_0$, its admissible stress $\sigma_0$ and geometry of the coil be prescribed.

Using (2.7) and (2.8), one determines the total current in the coil:

$$J_0 = a_c h_c \ln(K),$$

where $a_c$ is the value of coefficient $a$ of the Eq. (2.8) for the constant thickness Bitter coil.

The application of Eq. (2.6) leads to the radial dependence of magnetic flux density in the plane of plate as the function of radius $r$:

$$B_c(r) = \frac{\mu_0 J_0}{2r \ln K} \left( 1 - \frac{r}{K R_1} \right). \tag{3.1}$$

The magnetic flux density inside the inner radius of coil is equal to:

$$B_c(R_1) = \frac{\mu_0 J_0}{2 R_1 \ln K} \left( 1 - \frac{1}{K} \right).$$

Afterward we determine the stresses in Bitter plate of the constant thickness. The Eq. (2.2),(2.4) for the constant thickness reduce to:

$$-r^2 \frac{d^2 u}{dr^2} - r \frac{du}{dr} + u + F = 0 .$$

The distributed Lorentz force density per unit radius of the windings with constant thickness is the function of radius:

$$F = \left( K - \frac{r}{R_1} \right) f_c ,$$

$$f_c = \frac{\mu_0 J_0^2 \left( \nu^2 - 1 \right)}{2 E h_c K \ln^2 K}.$$

The general solution for radial displacement field reads:

$$u = f_c \left( \frac{r}{4 R_1} - K - \frac{r}{2 R_1} \ln r \right) + c_1 r + \frac{c_2}{r} . \tag{3.2}$$

The unknown constants $c_1$, $c_2$ are to be determined from the absence of radial pressure on the inner and outer surfaces of the coil:

$$c_1 = f_c \left[ 2(K^2 - 1)(1 + \nu)\ln R_1 + \left(K^2(3 + 2\ln K) + 1 - 4K\right)\nu + K^2 - 1 + 2K^2 \ln K \right]$$

$$c_2 = f_c R_1 \frac{K^2}{2(K^2 - 1)(1 - \nu)} \left(2\nu(1 - K) + (1 + \nu)\ln K\right) \tag{3.3}$$

With the expressions (3.2) - (3.3) the radial and circumferential stresses are determined to:

$$\sigma_r = \frac{E}{1 - \nu} c_1 - \frac{E}{1 + \nu}\frac{c_1}{r^2} + \frac{Ef_c}{r(\nu^2 - 1)R_1} \left[\nu(2\ln r + 4KR_1 - r) + r(1 + 2\ln r)\right],$$

$$\sigma_\theta = \frac{E}{1 - \nu} c_1 + \frac{E}{1 + \nu}\frac{c_1}{r^2} + \frac{Ef_c}{r(\nu^2 - 1)R_1} \left[2\ln r + 4KR_1 - r + \nu r(1 + 2\ln r)\right]. \tag{3.4}$$

With the formulas (3.4) the equivalent stress is evaluated according to Eq. (2.3). The equivalent stresses on the outer and inner surfaces of coil are:

$$\sigma_{eqv}(r = R_2) = \frac{\mu_0 J_0^2}{4h_c R_1} \frac{\left(-4K + 3 + 2\ln K + K^2\right)\nu + 1 + 2\ln K - K^2}{K(K - 1)(K + 1)\ln^2 K},$$

$$\sigma_{eqv}(r = R_1) = \frac{\mu_0 J_0^2}{4h_c R_1}$$

$$\frac{\left(-2K^3 + 3K^2 - 2K + 1 + 2K^2 \ln K\right)\nu - 2K^3 + K^2 + 2K - 1 + 2K^2 \ln K}{K(K - 1)(K + 1)\ln^2 K}. \tag{3.5}$$

The stress on the inner surface of the coil is higher, that the stress on the outer surface of the Bitter coil:

$$\sigma_{eqv}(r = R_1) > \sigma_{eqv}(r = R_2).$$

The condition that the equivalent stress equals to its ultimate admissible value:

$$\sigma_{eqv}(r = R_1) = \sigma_0 \tag{3.6}$$

is the wanted equation that is required for determination of constant thickness $h_c$. With this equation is possible to resolve the up till now unknown value $h_c$. Solving the equation (3.6) the thickness of the coil results to:

$$h_c = \frac{\mu_0 J_0^2}{4R_1 \sigma_0} \cdot$$

$$\frac{\left(-2K^3 + 3K^2 - 2K + 1 + 2K^2 \ln K\right)\nu - 2K^3 + K^2 + 2K - 1 + 2K^2 \ln K}{K(K - 1)(K + 1)\ln^2 K} \tag{3.7}$$

With this expression the calculation of the stress-limited design of Bitter coil with constant thickness is completed.

# 4. Static magnetic analysis of the Bitter coil of optimal thickness

At second consider the Bitter coil of the variable thickness $h(r)$. Our task is to determine the optimal radial distribution of thickness. We seek the thickness distribution that leads to the largest possible magnetic field strength under the consition of admissible stress due to Lorenz forces.

The total current in the coil $J_0$, its admissible stress $\sigma_0$ and inside and outside radii of the coil of the optimal and constant thickness Bitter coils are assumed to be equal.

For optimality, the circumferential stress over the coil is assumed to be constant and equal to its upmost admissible value:

$$\sigma_\theta = \frac{E}{1-v^2}\left[v\frac{du}{dr}+\frac{u}{r}\right]=-\sigma_0,$$ (4.1)

From the Eq. (4.1) the radial displacement could be immedeatly determined. The general solution of Eq. (4.1) reads to radial displacement and the corresponding radial component of stress :

$$u = \frac{(1-v)\sigma_0 r}{E}+c_0 r^{-\frac{1}{v}}, \qquad \sigma_r = -\sigma_0 - \frac{c_0 E}{v}r^{-1-\frac{1}{v}}.$$ (4.2)

From the condition (3.6) follows, that the integration constant vanishes:

$$c_0 = 0.$$

Substitution of Eq. (4.2) in Eq. (2.4) leads to the equation for the optimal thickness $h_o(r)$:

$$\frac{d[h_o(r)\sigma_0]}{dr}+B_o(r)h_o(r)j(r)=0.$$ (4.3)

However, in the Eq. (4.3) the magnetic flux density of the optimally shaped coil $B_o(r)$ depends on $h_o$ through the integral expressions (2.6) and (2.8). Because of this circumstance the Eq. (4.3) could not be solved immediately in terms of thickness $h_o$ as the ordinary differential equation. Namely, the Eq. (4.3) is in fact an integral equation.

This difficulty could be easily avoided. Namely, instead of solving the integral equation in terms of thickness $h_o$, it is conveniently to deal at first with differential equation with respect to $B_o(r)$. For this purpose the function $h_o$ eliminates using the expressions (2.6) and (2.8):

$$h_o(r)=-\frac{2r^2}{\mu_0 a}\frac{dB_o}{dr}.$$ (4.4)

Substitution of Eq. (4.4) into (4.3) leads to an ordinary differential equation in terms of the magnetic flux density of the optimally shaped coil $B_o(r)$:

$$r\frac{d^2 B_o}{dr^2} + 2\frac{dB_o}{dr} - \frac{a_o}{\sigma_0}B_o(r)\frac{dB_o}{dr} = 0 \ . \tag{4.5}$$

The solution of Eq. (4.5) with two new integration constants $c_1, c_2$ reads:

$$B_o(r) = -\frac{\sigma_0}{a_o}\left[1 + \sqrt{2c_1 a_o \sigma_0 + \sigma_0{}^2}\ \tanh\left(\frac{\sqrt{2c_1 a_o \sigma_0 + \sigma_0{}^2}}{2\sigma_0}\ln\frac{r}{c_2}\right)\right]. \tag{4.6}$$

With the substitution

$$c_1 = 0, \qquad c_2 = R_1$$

the Eq. (4.6) reduces to:

$$B_o(r) = \frac{2\sigma_0 R_1}{(R_1 + r)a_o}\ . \tag{4.7}$$

Using Eq. (4.4) the chosen optimal thickness of Bitter coil appears as the function of radius:

$$h_o(r) = \frac{4\sigma_0 R_1}{\mu_0 a_o{}^2}\frac{r^2}{(R_1 + r)^2}\ , \tag{4.8}$$

$$\frac{h_o(R_2)}{h_o(R_1)} = \frac{4\mathrm{K}^2}{(\mathrm{K}+1)^2}\ . \tag{4.9}$$

With the expression for optimal thickness the total current through one Bitter coil with (2.8) reads as:

$$J_0 = \int_{R_1}^{R_2} h_o(r)j(r)dr = \Phi(\mathrm{K})\frac{4\sigma_0 R_1}{a_o \mu_0}\ ,$$

$$\Phi(\mathrm{K}) = -\ln(1 + \mathrm{K}) + \ln 2 + \frac{1}{2}\frac{\mathrm{K}-1}{1+\mathrm{K}}\ . \tag{4.10}$$

Substituting of (4.7), (4.8) into (4.10) leads to the expression for the total current that depends on a single constant $a_o$. The constant $a_o$ concludes to:

$$a_o = \frac{4\sigma_0 R_1}{J_0 \mu_0}\Phi(\mathrm{K})\ .$$

With this value for constant $a_o$ the explicit formula for optimal thnckness results:

$$h_o(r) = \frac{1}{\Phi^2(\mathrm{K})}\frac{r^2}{R_1(R_1 + r)}\frac{J_0{}^2 \mu_0}{4\sigma_0}\ . \tag{4.11}$$

The optimal thnckness (4.11) depends solely on total current $J_0$, admissible stress $\sigma_0$ and inner and outer radii of plate. Correspondingly, for the optimal thickness the magnetic induction of optimal solenoid evaluates as:

$$B_o\left(R_1\right)=\frac{J_0\mu_0}{4\Phi(\mathrm{K})}\frac{1}{R_1}\,.$$

(4.12)

The material volume of optimal solenoid with (2.8) is the explicit function of prescribed parameters:

$$V=2\pi\int_{R}^{R_t}h_o(r)r\,dr=\frac{J_0{}^2\pi\mu_0 R_1}{\sigma_0}\frac{\Psi(\mathrm{K})}{\Phi^2(\mathrm{K})},$$

$$\Psi(\mathrm{K})=-6\ln(1+\mathrm{K})+6\ln 2+\frac{1-\mathrm{K}}{1+\mathrm{K}}\left(\mathrm{K}^2-2\mathrm{K}-4\right)$$

The ratio of the magnetic flux densities of the optimal to constant thickness coils is equal to:

$$\kappa_1=\frac{B_o}{B_c}=\frac{\mathrm{K}\ln\mathrm{K}}{2(\mathrm{K}-1)\Phi(\mathrm{K})}\,.$$

(4.13)

The distance between the narrowed Bitter plates at the inner diameter could be used for cooling fluid channels (**Fig.1**). The cross-section of cooling fluid channels increases to the inner diameter of solenoid together with the current density. This situation optimizes the thermal flux in the most intensively heated region. The cooling process of the optimal solenoid with the radial distribution of the conducting plate thickness is also considerably enhanced.

# 5. Comparison of the constant and optimal thickness Bitter coils. Example 1.

In this section the comparison of the actual design with the constant and the design with optimal Bitter coil plates is provided for Fawley Power Station unit (Caldwell, 1982).

The total current for active coils $J_0$ and the total thickness of active coils $h(R_1)$ is calculated for the following values of the parameters:

$\sigma_0 = 3.7 \cdot 10^7 \, Pa, \qquad J_0 = 3 \cdot 10^6 \, A, \quad R_1 = 0.55m, \quad R_2 = 1.29m$

$h_o(R_1) = 0.3739 \, m, \qquad h_o(R_2) = 0.7351 \, m, \qquad h_c = 0.173 \, m.$

The magnetic flux densities of the optimal and constant thickness Bitter plates are

$B_o = 5.4683 \, \text{T}, \qquad U_o = B_o^{\,2} / (2\mu_0) = 0.1189 \cdot 10^8 \, Pa,$

$B_c = 2.3061 \, \text{T}, \qquad U_c = B_c^{\,2} / (2\mu_0) = 0.2116 \cdot 10^7 \, Pa.$

The following plots expose the main physical quantities of the reference and new optimal designs of solenoids. All plots with the red color correspond to reference design with the constant thickness of Bitter plates. The plots colored blue depict the optimal, constantly stressed-design of Bitter plates.

Magnetic flux density of the optimal and the constant thickness coils are shown on the **Fig. 2**. On the **Fig. 3** the equivalent stresses of the optimal and the constant thickness coils are visualized. For the constant thickness coil the stress varies. Correspondingly, the stress is constant of the optimal thickness distribution of Bitter plate. The current densities of the optimal and the constant thickness coils are imagined on **Fig. 4**. The corresponding plots of the optimal and the constant thickness coils depicts the **Fig. 5**.

# 6. Comparison of the constant and optimal thickness Bitter coils. Example 2

The comparison of the designs with the constant and optimal Bitter coil plates is provided for the magnet of RUTHERFORD HIGH ENERGY LAB (Caldwell, 1982).

The total current for active coils $J_0$ and the total thickness of active coils $h(R_1)$

$$\sigma_0 = 6.9 \cdot 10^7 \, Pa, \qquad J_0 = 1.02 \cdot 10^7 A, \qquad R_1 = 0.95 m, \quad R_2 = 1.70 m,$$
$$h_o(R_1) = 3.92 \, m, \qquad h_o(R_2) = 6.45 \, m, \qquad h_c = 0.99 \, m.$$

The magnetic flux densities of the optimal and constant thickness Bitter plates are

$$B_o = 17.64 \, \text{T}, \qquad U_o = {B_o}^2 / (2\mu_0) = 0.123 \cdot 10^9 \, Pa.$$
$$B_c = 5.1145 \, \text{T}, \qquad U_c = {B_c}^2 / (2\mu_0) = 0.104 \cdot 10^8 \, Pa.$$

Magnetic flux density of the optimal and the constant thickness coils are exposed on the **Fig. 6**. On the **Fig. 7** the equivalent stresses of the optimal and the constant thickness coils are shown. Once again, for the constant thickness coil the stress varies. The stress is constant of the optimal thickness distribution of Bitter plate. The **Fig.8** demonstrates the current densities of the optimal and the constant thickness coils. The **Fig. 9** represents the thicknesses of the optimal and the constant coils.

## 7. Conclusion

The new accurate magnetic force formulas for the system of the Bitter disk with the radially varying thickness and current are derived and presented in this paper. The most important result demonstrates the considerable increase of the magnetic field density using the optimal profiling of the thickness of Bitter plates. The distance between the narrowed Bitter plates at the inner diameter could be used for cooling fluid channels. The cross-section of cooling fluid channels increases to the inner diameter of solenoid together with the current density. This situation optimizes the thermal flux in the most intensively heated region and improves the cooling process of the optimal solenoid with radially profile of the plate thickness.

 All expressions are obtained in the close form. Also we gave in this paper the evaluation of optimization effects in the actual Bitter solenoids. The presented method can be used to design of solenoids and calculation of important electrical quantities as the magnetic force and magnetic field strength.

# Symbols

| | |
|---|---|
| $R_1, R_2, R_1 \le r \le R_2$ | Inside and outside radii of the Bitter coil |
| $R_2/R_1 = K > 1$ | Radius ratio |
| $u$ , | Radial component of displacement |
| $\varepsilon_r, \varepsilon_\theta$ | Radial and circumferential deformation |
| $\sigma_r, \sigma_\theta$ | Radial and circumferential stress |
| $\sigma_{eqv}$ | Equivalent stress |
| $h$ | Thickness of the Bitter coil |
| $F$ | The distributed Lorentz force density per unit radius |
| $B$ | The magnetic flux density (field intensity) |
| $j$ | The current density pro unit area |
| $f_0$ | The constant |
| $V$ | The material volume of solenoid |
| $h_c$ | The Bitter coil of the constant thickness |
| $a_c$ | The value of coefficient $a$ for of the constant thickness |
| $B_c$ | The magnetic flux density for the constant thickness coil |
| $B_o$ | The magnetic flux density of the optimally shaped coil |
| $U_o = B_o^{\ 2}/(2\mu_0)$ | Pressure of magnet field |
| $U_c = B_c^{\ 2}/(2\mu_0)$ | |

## 8. References

Herlach F., Jones H. (1994) Magnets, Encyclopedia of Applied Physics, Ed. G.L. Trigg, Wiley-VCH Verlag, Vol. 9, p. 245-260.

Knoepfel H.E. (2000) Magnetic Fields, John Wiley & Sons, N.Y.

Kratz R., Wyder P. (2013) Principles of Pulsed Magnet Design, Springer Science & Business Media, Berlin, Heidelberg

Mulhall B.E., Prothero D.H. (1973) Mechanical stresses in solenoid coils, J. Phys. D: Appl. Phys., Vol. 6, 1973.

Caldwell J. (1980) Modelling of the stress distribution in superconducting magnet windings, Appl. Math. Modelling, 1980, Vol 4, p.228

Caldwell J. (1982) Electromagnetic forces in high field magnet coils, Appt. Math. Modelling, Vol. 6, June, p. 157

Caldwell J. (1982) The circumferential stress in large high field magnet coils, Appl. Math. Modelling, 1982, Vol 6, February, p.67

Markiewicz W.D., Vaghar M.R., Dixon I.R., Garmestani H. (1999) Generalized plane strain analysis of superconducting solenoids, J. Appl. Phys, Vol. 86, No. 12

Motokawa M., Nojiri H., Tokunaga Y. (1989) An idea for the easy construction of a high field magnet, Physica B: Condensed Matter, Vol. 155, Issues 1–3, p. 96-99

## Optimal Bitter Coil Solenoid

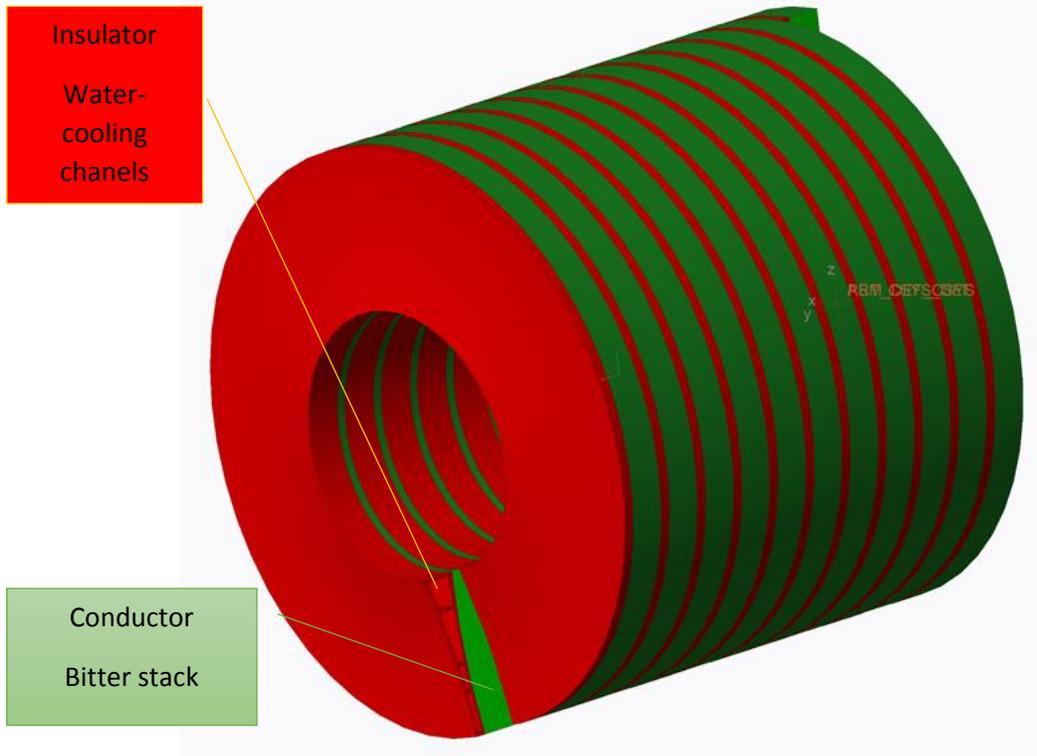

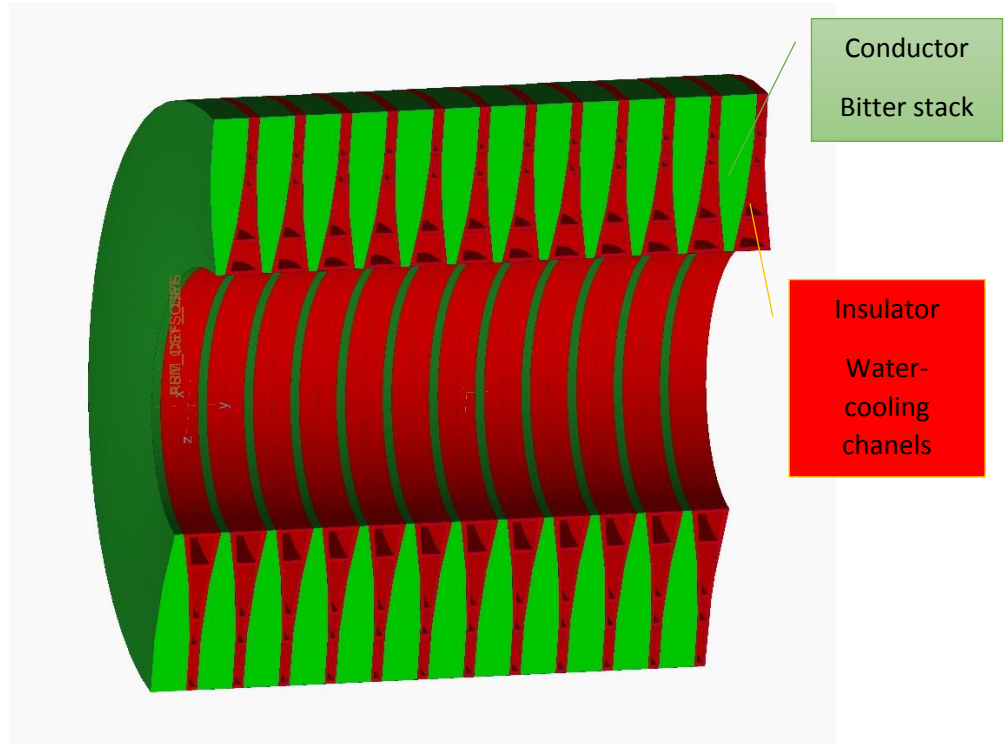

**Fig. 1 Design of the core element of the optimal Bitter solenoid**

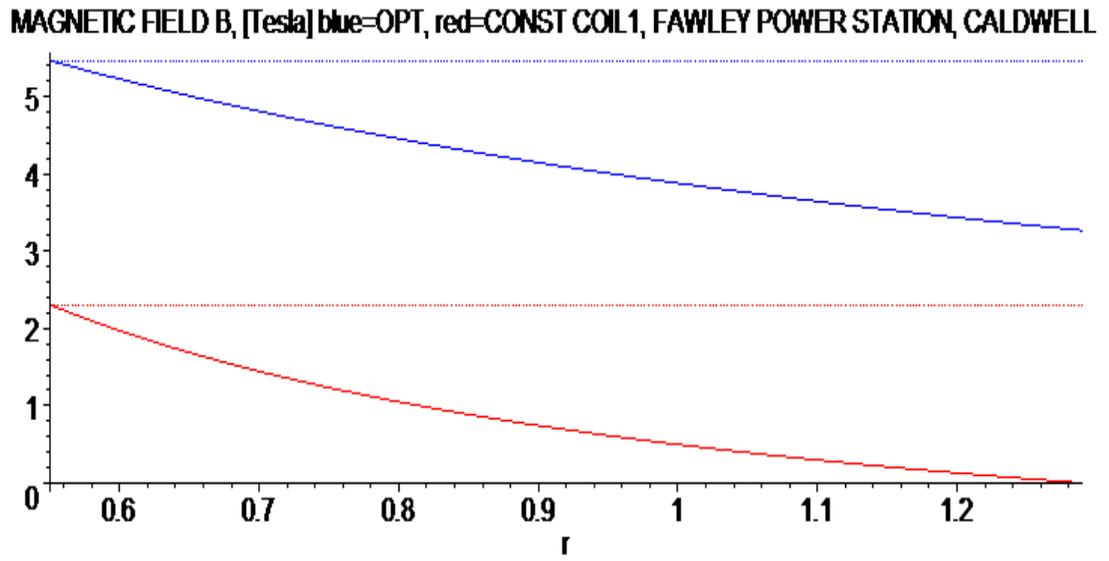

Fig. 2 Magnetic flux density of the optimal and the constant thickness coils

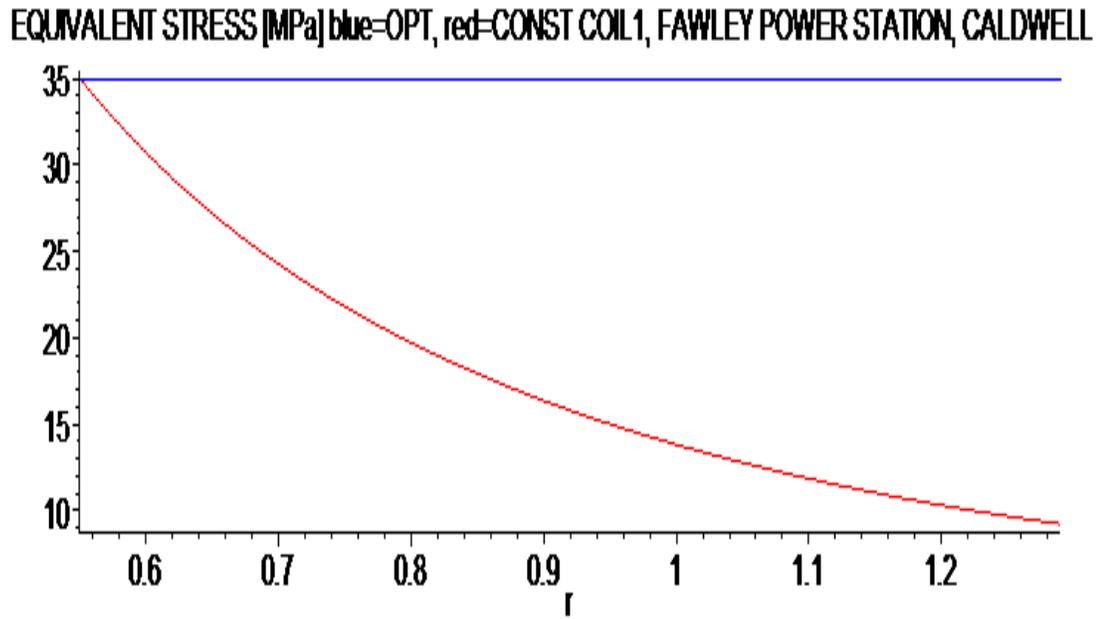

**Fig. 3 Equivalent stresses of the optimal and the constant thickness coils**

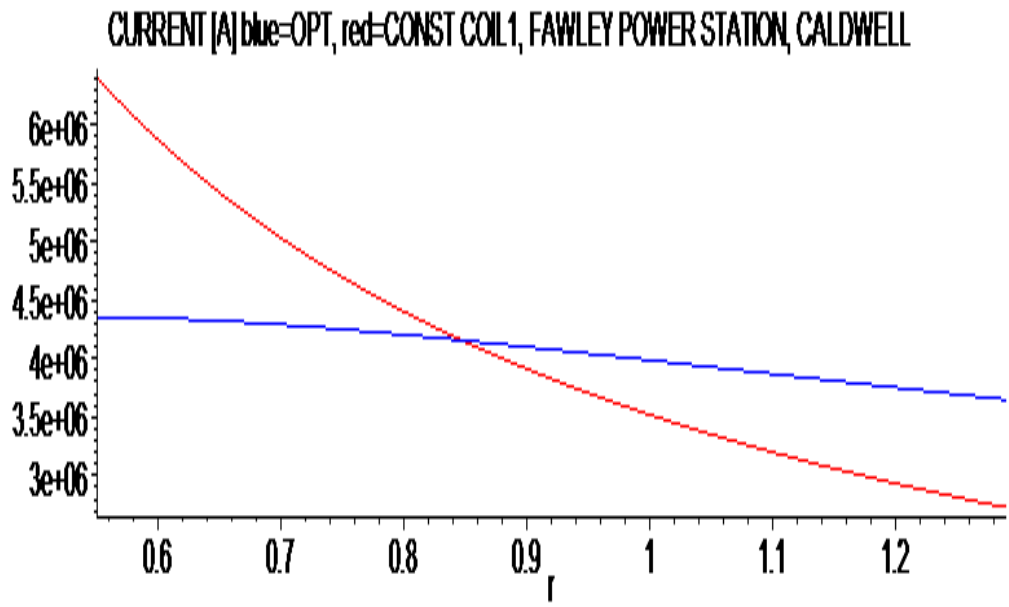

**Fig. 4 Current densities of the optimal and the constant thickness coils**

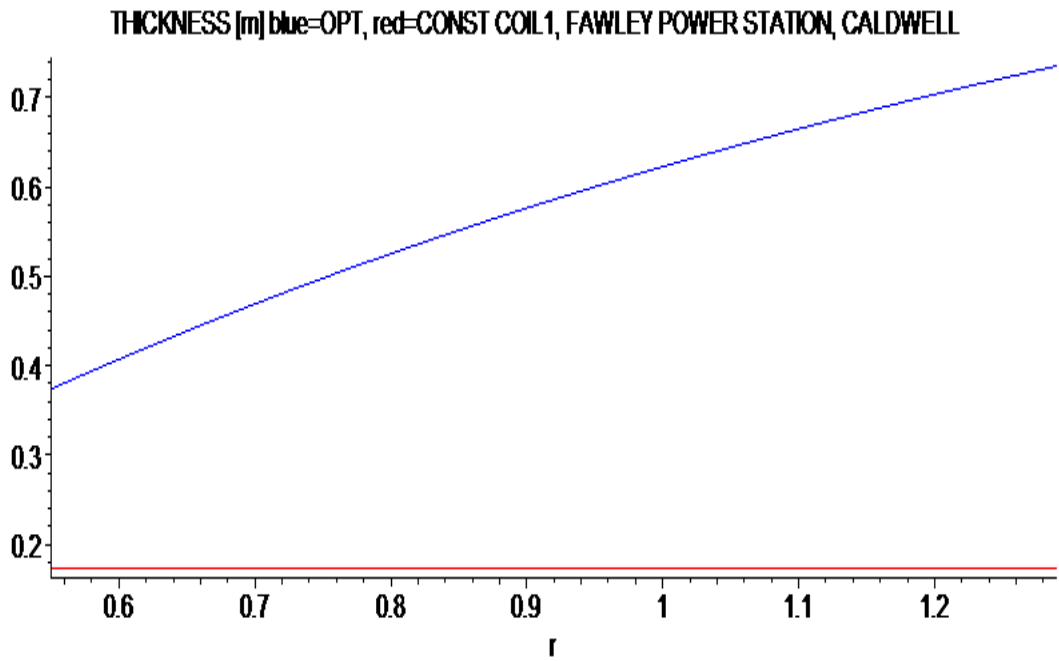

**Fig. 5 Thicknesses of the optimal and the constant thickness coils**

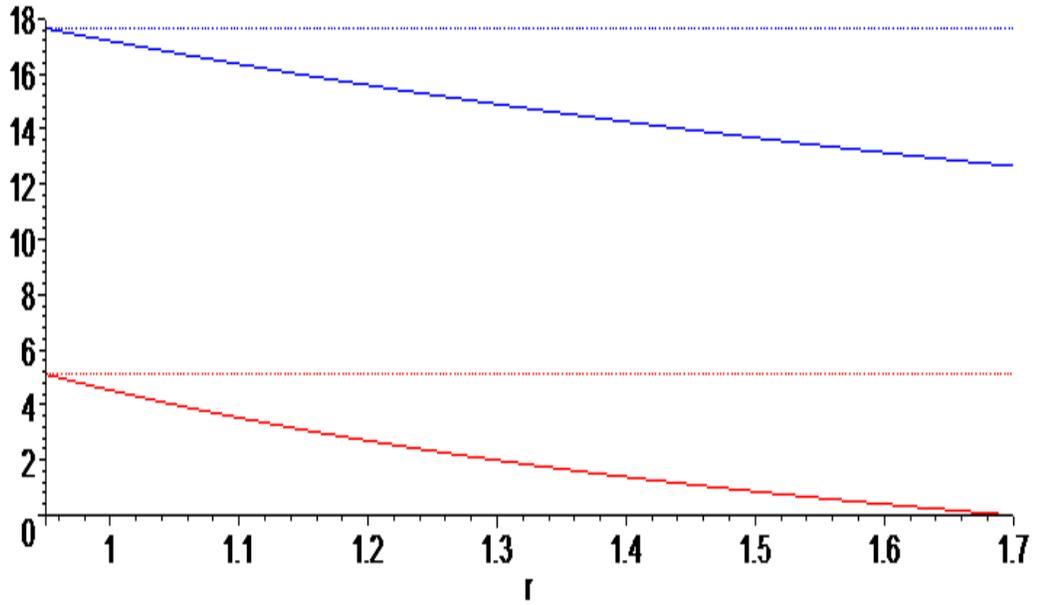

**Fig. 6 Magnetic flux density of the optimal and the constant thickness coils**

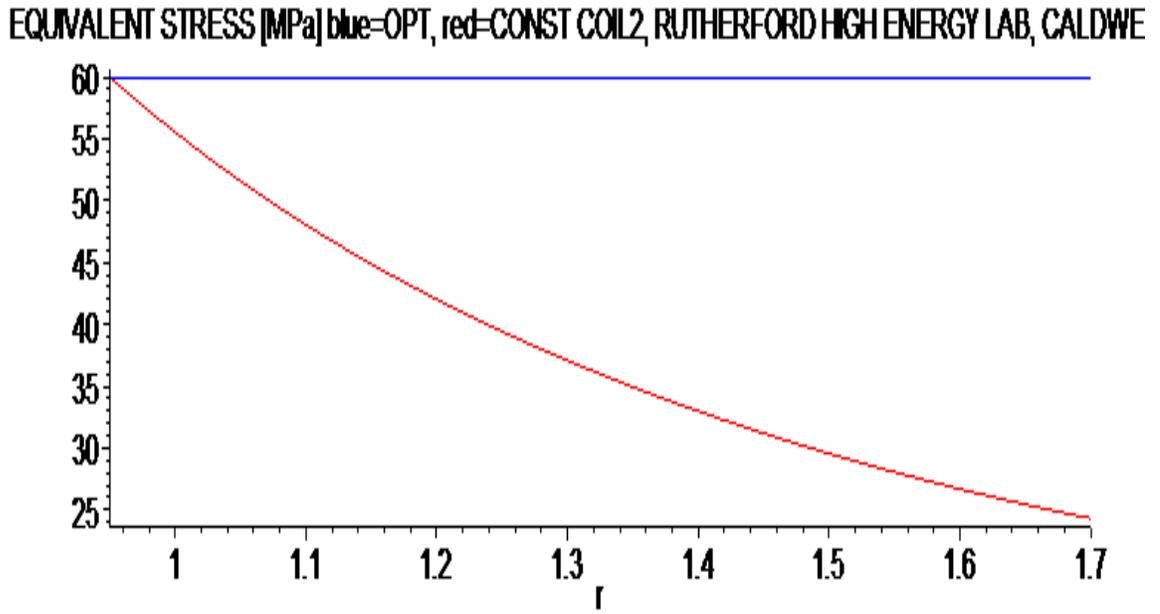

**Fig. 7 Equivalent stresses of the optimal and the constant thickness coils**

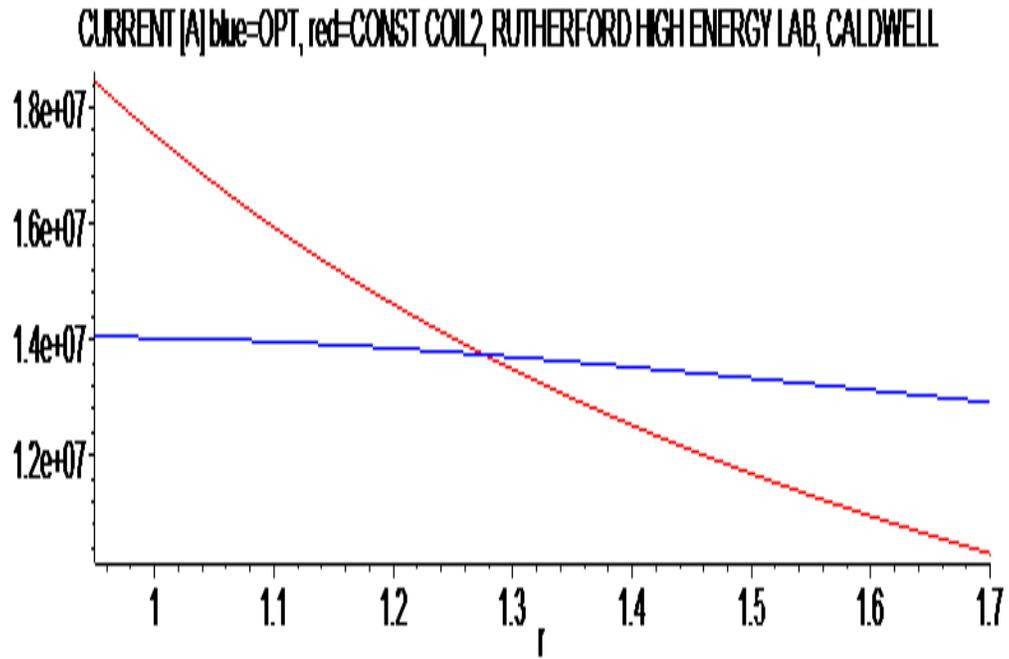

**Fig. 8 Current densities of the optimal and the constant thickness coils**

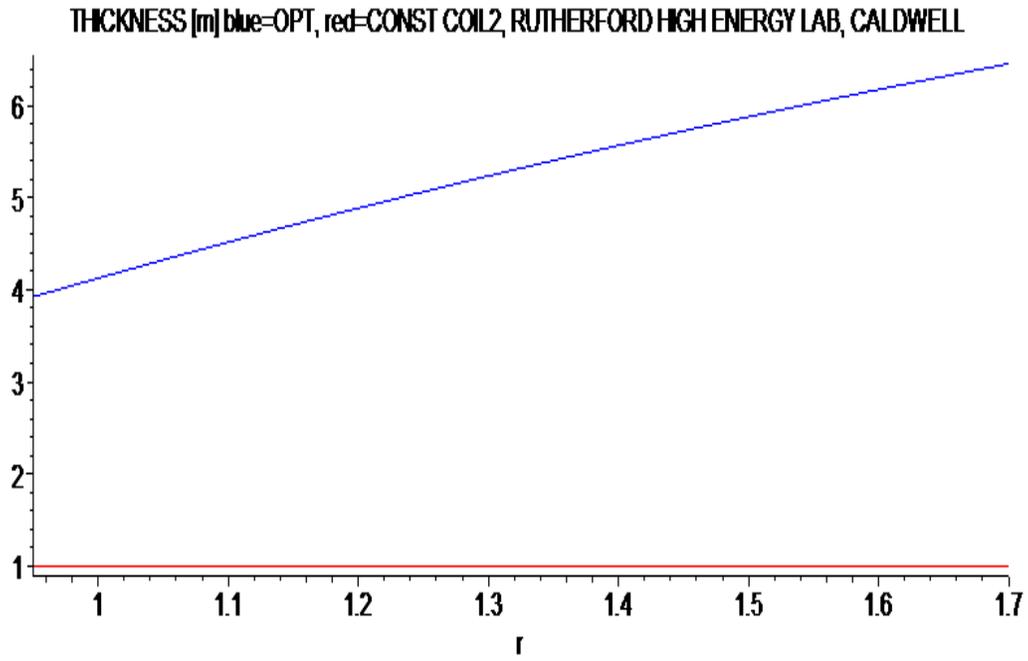

THICKNESS [m] blue=OPT, red=CONST COIL2, RUTHERFORD HIGH ENERGY LAB, CALDWELL

**Fig. 9 Thicknesses of the optimal and the constant thickness coils**